\documentclass[a4paper,american]{scrartcl}
\usepackage{tikz-cd}
\usepackage{verbatim}
\usepackage{lmodern}
\usepackage[utf8]{inputenc}
\usepackage[T1]{fontenc}
\usepackage{color}
\usepackage{todonotes}
\usepackage{babel}
\usepackage{amsfonts}
\usepackage{amsmath}
\usepackage{amssymb}
\usepackage{setspace}
\usepackage[authoryear]{natbib}
\usepackage[font=small]{quoting}
\usepackage{latexsym}
\usepackage[bookmarks=false,pdfborder={0 0 0},colorlinks=false]{hyperref}
\usepackage[normalem]{ulem}
\usepackage{enumerate}
\date{}

\makeatletter

\newcounter{contatore}
\newtheorem{ax}[contatore]{Axiom}

\newcounter{contatore1}
\newtheorem{ed}[contatore1]{Definition}

\newcounter{contatore2}
\makeatother

\begin{document}

\title{Relationalism about mechanics based on a minimalist ontology of matter}

\author{Antonio Vassallo%
\thanks{Université de Lausanne, Faculté des lettres, Section de philosophie,
1015 Lausanne, Switzerland. E-mail:
\protect\href{mailto:Antonio.Vassallo@unil.ch}{Antonio.Vassallo@unil.ch}}%
, Dirk-Andr\'e Deckert%
\thanks{Ludwig-Maximilians-Universität München, Mathematisches Institut,
Theresienstrasse 39, 80333 München, Germany. E-mail: \protect\href{mailto:deckert@math.lmu.de}{deckert@math.lmu.de}%
}, Michael Esfeld%
\thanks{Université de Lausanne, Faculté des lettres, Section de philosophie,
1015 Lausanne, Switzerland. E-mail: \protect\href{mailto:Michael-Andreas.Esfeld@unil.ch}{Michael-Andreas.Esfeld@unil.ch}}
}

\maketitle
\vspace{-1.5cm}

\begin{center} \emph{Accepted for publication in European Journal for Philosophy of Science}
    \end{center}
\begin{abstract} 
    
    %\medskip

    This paper elaborates on relationalism about space and time as motivated by a minimalist ontology of the physical world: there are only matter points that are individuated by the distance relations among them, with these relations changing. We assess two strategies to combine this ontology with physics, using classical mechanics as example: the Humean strategy adopts the standard, non-relationalist physical theories as they stand and interprets their formal apparatus as the means of bookkeeping of the change of the distance relations instead of committing us to additional elements of the ontology. The alternative theory strategy seeks to combine the relationalist ontology with a relationalist physical theory that reproduces the predictions of the standard theory in the domain where these are empirically tested. We show that, as things stand, this strategy cannot be accomplished without compromising a minimalist relationalist ontology.\medskip

\end{abstract}

\tableofcontents{}

\section{From atomism to relationalism about space and time}
\label{sec:motivation}

Atomism, going back to the pre-Socratic philosophers Leucippus and Democritus and turned into a precise physical theory by Newton, is the most successful paradigm in both classical physics and traditional natural philosophy. On the one hand, it is a proposal for a fundamental ontology that is most parsimonious and most general, applying to everything in the universe. On the other hand, it offers a clear and simple explanation of the realm of our
experience. Macroscopic objects are composed of fundamental, indivisible particles. All the differences between the macroscopic objects -- at a time as well as in time -- are accounted for in
terms of the spatial configuration of these particles and its change.

However, there is no straightforward answer to the question of what are the atoms. Both Democritus and Newton adopt the view of the atoms being inserted into an absolute space. Consequently, they are committed to a dualism between on the one hand space and on the other hand matter in the guise of atoms filling space. But what is it that fills space? In other words, what makes up the difference between a location in space being occupied by an atom and its being empty? Physical properties taken to characterize matter -- such as mass, charge, or spin, etc. -- are introduced in physical theories in terms of their causal role for the evolution of the configuration of matter. Consequently, invoking these properties cannot answer the question of what it is that evolves in space as described or prescribed by these properties (see \citealp{Blackburn:1990aa}).

To put it differently, the parameters that figure in the equations of a physical theory \emph{presuppose} a spatial configuration of matter to which they are applied. This is particularly evident in the case of the quantum state as represented by the wave function, which is defined on configuration space -- that is, the mathematical space each of whose points represents a possible configuration of matter in physical space. Hence, the quantum state presupposes a configuration of objects in physical space to which it is applied. But also in classical mechanics, parameters such as mass and charge presuppose objects given in terms of their spatial location to which they are applied.

If one reacts to this situation by taking the objects in space to be bare substrata (cf. \citealp{Locke:1690aa}, book II, chapter XXIII, § 2), one runs into the problem that a bare substratum or primitive stuff-essence of matter is mysterious. The same goes for a primitive thisness (haecceity) of the material objects. However, this impasse of not being able to come up with a characterization of matter that stands up to scrutiny arises only if one accepts a dualism of an absolute space and matter as that what fills space. If one abandons this dualism and conceives atomism in terms of relationalism about space, then the spatial relations are available to answer the question of what the atoms are. This is the idea that we shall pursue in this paper, making use of the Cartesian conception of matter in terms of (spatial) extension and the stance of ontic structural realism according to which objects are individuated by the relations in which they stand -- i.e. distance relations in our case.

In other words, our claim is that atomism, if set out in terms of point particles being inserted into an absolute space, fails to achieve the aim of being a parsimonious ontology. The consequence of this failure is that atomism, thus conceived, is unable to formulate a cogent answer to the question of what the atoms are. To meet the requirement of parsimony, one has to abandon points of space and retain only point particles (matter points), with these point particles standing in distance relations that individuate them. Atomism hence motivates relationalism about space and time instead of being tied to the commitment of an absolute space and time.

Furthermore, relationalism is motivated by the fact that the commitment to absolute space and time introduces a surplus structure that is not needed to account for the empirical evidence. Thus, Leibniz points out in his famous objections to Newton's substantivalism that there are many different possibilities to place or to transform the whole configuration of matter in an absolute space that leave the spatial relations among the material objects unchanged so that there is no physical difference between them (see notably Leibniz' third letter, §§ 5-6, and fourth letter, § 15, in \citealp{Leibniz:1890aa}, pp. 363-364, 373-374, English translation \citealp{Leibniz:2000aa}). However, Leibniz' objection does not apply to all forms of substantivalism in classical physics, not to mention relativity physics. For instance, it can be circumvented in neo-Newtonian space-time (see e.g. \citealp{Maudlin:1993aa}, p. 192; see furthermore \citealp{Pooley:2013aa}, for a recent and comprehensive overview of the substantivalism/relationalism debate).

Nonetheless, the general objection of introducing a commitment to surplus structure hits any form of space-time substantivalism: assume, as is well supported by all the available physical evidence, that the configuration of matter consists in finitely many discrete objects, such as point particles. If that configuration is embedded in an absolute space, then that space will stretch out to infinity, unless an arbitrary boundary is imposed (at least in a classical setting, since in general relativity the global matter distribution might determine a compact geometry); in any case, it will stretch out far beyond the actual particle configuration. However, all the experimental evidence is one of relative particle positions and change of particle positions, that is, motion. Thus, space is needed in physics only to describe the configuration of matter and, notably, the change in that configuration. Consequently, subscribing to the existence of an absolute space in which that configuration is embedded amounts to inflating the ontology.      

Against this background, our claim is that in order to accomplish the task of elaborating on a parsimonious ontology of the physical world -- at least as far as the setting of classical, pre-relativistic physics is concerned --, only the following two axioms are required:

 \begin{ax}\label{a1}
    There are distance relations that individuate objects, namely matter points.
  \end{ax}

  \begin{ax}\label{a2}
     The matter points are permanent, with the distances between them changing.
  \end{ax} 

We submit that these two axioms are necessary and minimally sufficient to formulate an ontology of the physical world in the context of classical, pre-relativistic physics that is empirically adequate, given that all the empirical evidence comes down to relative particle positions and change of these positions.

Why should one single out the distance relations? If there is a plurality of objects, there has to be a certain type of relations in virtue of which these objects make up a configuration that then is the world. Generally speaking, one can conceive different types of relations making up different sorts of worlds. For instance, one may imagine thinking relations that individuate mental substances making up a world of minds, etc. Lewis's hypothetical basic relations of like-chargedness and opposite-chargedness, by contrast, would not pass the test, since, as Lewis notes himself, these relations fail to individuate the objects that stand in them as soon as there are at least three objects (\citealp{Lewis:1986ab}, p. 77).

When it comes to the natural world, the issue are relations that qualify as providing for extension. That is the reason to single out distance relations. In a future theory of quantum gravity, these relations may be conceived in a different manner than in our current and past physical theories. Nonetheless, we submit that relations providing for extension -- namely distances -- are indispensable for an ontology of the natural world that is to be empirically adequate. Change in these relations then is sufficient to obtain empirical adequacy. That is the reason to pose the two above mentioned axioms, and only these two ones. Accordingly, distances individuating point-objects that then are matter points and change of these distances are the primitives of a minimalist ontology of the physical world, again at least as far as classical, pre-relativistic physics is concerned.

To convey what axiom \ref{a1} means, we have to choose a representation. Let us consider a universe consisting of a finite number of $N\in\mathbb{N}$ matter points. Taking the number of matter points to be finite is sufficient for empirical adequacy and will make the following discussion much easier. In order to obtain an ontology that is necessary and minimally sufficient for empirical adequacy, we assume that the set $\Omega$ of all possible configurations of distance relations between $N\in\mathbb{N}$ matter points can be represented as follows:
\begin{ed}\label{def:dist}
    Let $\mathcal M=\{1,2,\dots,N\}$ and $\mathcal E=\{(i,j)\,|\,i,j\in\mathcal
    M, i\neq j\}$. The set $\Omega$ comprises
    elements $\Delta=(\Delta_{ij})_{(i,j)\in\mathcal E}$ that can be represented by numerical assignments fulfilling the following requirements:
    \begin{enumerate}[(i)]
        \item\label{one} $\Delta=(\Delta_{ij})_{(i,j)\in\mathcal E}$ is a $\frac{N}{2}(N-1)$-tuple of positive values
            $\Delta_{ij}\in\mathbb R^{+}$ for each $(i,j)\in\mathcal E$.
        \item\label{two} For all $(i,j)\in\mathcal E$ one has $\Delta_{ij}=\Delta_{ji}$.
        %\item\label{zero} For all $(i,j)\in\mathcal E$ one has $\Delta_{ij}=0$ if and only if $i=j$.
        \item\label{three} For all $i,j,k\in \mathcal M$, it is the case that $\Delta_{ij}\leq\Delta_{ik}+\Delta_{kj}$.
        \item\label{four} For all $i \neq j$ there is a $k \neq i,j$ such that $\Delta_{ik} \neq \Delta_{jk}$.
        %For all  $i\neq j\in \mathcal M$, there exists at least one $k\neq i,j$ such that $\Delta_{ik}\neq\Delta_{jk}$.
    \end{enumerate}
\end{ed}

Due to requirements (\ref{one}) and (\ref{two}), the distance relation is irreflexive, symmetric, and connex. Requirement (\ref{three}) is the triangle inequality in virtue of which the relation is a distance relation. Requirement (\ref{four}) states that the distance relations individuate the matter points: if matter point $i$ is distinct from matter point $j$, there exists at least one other matter point $k$ such that matter points $i$ and $j$ are distinguished by their relation to $k$. Formulating these requirements in terms of numerical assignments is a means to express the features in virtue of which a relation is a distance relation. However, the numerical assignments do not belong to the ontology, let alone the notion of absolute scale that comes with them; they are just introduced for representational purposes. Consequently, the fact that the values assignable to $\Delta_{ij}$ are real numbers does not smuggle in any infinity in this ontology: there is a finite number of $N$ matter points and, hence, finitely many distance relations.
 
Nonetheless, by introducing a labelling $\mathcal M$ of the matter points, this definition can be taken to suggest that their numerical plurality is a primitive fact. But
this is just an artifact of the set-theoretical language. The elements of a set
$\mathcal{M}$ \emph{qua} set-theoretical objects have to be numerically distinct
for $\mathcal{M}$ to be a well-defined set of $N$ objects. However, the
referents of this formalism -- the matter points \emph{qua} physical objects --
are individuated by the distance relations given by $\Delta$, 
so that these relations account for their numerical plurality.

To emphasize the indistinguishability of the matter
points in the formalism, it is possible to make the above definition of
$\Omega$ independent of the labelling by introducing the following equivalence
relation:

\begin{ed}
    Take $\Delta,\Delta'\in\Omega$, and consider $\mathbb S_N$ as the set of all possible permutations of elements of $\mathcal M$. We define $\Delta\simeq\Delta'$ if and only if
    there is a permutation $\sigma\in \mathbb S_N$ such that for all
    $(i,j)\in\mathcal E$ it is the case that $\Delta'_{ij}=\Delta_{\sigma(i)\sigma(j)}$.
\end{ed}
The set
\begin{align}
    \widetilde \Omega = \Omega / {\simeq} := \left\{ [\Delta]_{\simeq} \, \big| \,
    \Delta \in \Omega \right\},
    \qquad
    [\Delta]_{\simeq} = \{\Delta'\in\Omega\,|\,\Delta'\simeq\Delta\}
\end{align}
then comprises all possible configurations of distance relations independently of
a labelling of the matter points. 

One way to envision an element of $[\Delta]_{\simeq}\in \widetilde \Omega$ is by
a representative $\Delta\in\Omega$ that can be viewed as a coloured graph
$G(\Delta)=(\mathcal M,\mathcal E,\Delta)$ in which $\mathcal M$ are the nodes,
$\mathcal E$ are the edges, and to each edge $(i,j)\in\mathcal E$ the colour
$\Delta_{ij}$ is attached. Also the graphs $G(\Delta)$ can be made
label-independent by considering the equivalence classes
$[G(\Delta)]_{\simeq}=\{G(\Delta')\,|\, \Delta'\simeq\Delta\}$ and treating $G(\Delta)$ only
as the corresponding representative of the class.

This ontology follows Leibniz' relationalism about space. According to Leibniz, distances make up the order of what coexists (third letter, § 4, in \citealp{Leibniz:1890aa}, p. 363). Distances are able to distinguish objects, thus respecting Leibniz' principle of the identity of indiscernibles. Hence, in virtue of these relations, there is a configuration of matter points that is constituted through variation in the distance relations that connect the matter points and that make it that these are matter points (in contrast to, say, mind points that are individuated by thinking relations). This ontology furthermore accounts for the impenetrability of matter without having to invoke a notion of mass: for any two matter points to overlap it would have to be the case that there is no distance between them.

Consequently, matter is structurally individuated, namely by the distances among the material objects. As the literature on ontic structural realism has made clear, structures in the sense of concrete
physical relations -- such as distances -- can individuate physical objects (see e.g. \citealp{Ladyman:2007aa}). Indeed, structures in this sense can do exactly the same what properties are supposed to do: if one holds that objects are bundles of properties, then the corresponding view is what is known as \emph{radical} ontic structural realism, namely the view that objects are constituted by relations, being the nodes in a network of relations (see \citealp{Ladyman:2007b}, chapters 2 and 3, and \citealp{French:2014aa}, chapters 5-7). If one thinks that there are underlying substances that instantiate properties, then the corresponding view is what is known as \emph{moderate} ontic structural realism, namely the view that objects and relations are on a par, being mutually ontologically dependent: relations require relata in which they stand, but all there is to the relata is given by the relations that obtain among them (see \citealp{Esfeld:2011aa}). In any case, the fundamental objects do not have an intrinsic nature, but a relational one.

In order to obtain the result that the distance relations individuate the matter points and thus distinguish them, one has to require that these relations establish what is known as absolute discernibility in today's literature: each of the matter points distinguishes itself from all the other ones by at least one distance relation that it bears to another matter point. What is known as weak discernibility in today's literature would not be enough, since weak discernibility does not avoid having to endorse a given numerical plurality of objects (\emph{contra} our parsimony requirement): for weak discernibility to be satisfied, it is sufficient that objects stand in an irreflexive relation, without there being anything that distinguishes one object from the other ones. Hence, weak discernibility indicates that there is a numerical plurality of objects, but is too weak to individuate the objects (the debate about weak discernibility goes back to \citealp{Saunders:2006aa}; as regards the distance relations, see the exchange between \citealp{Wuthrich:2009aa} and \citealp{Muller:2011aa}).

By way of consequence, any model of a theory that qualifies as genuinely relationalist by these standards has to include at least three matter points and has to comply with requirement (\ref{four}), thus ruling out notably symmetrical configurations. However, this is no objectionable restriction: having empirical adequacy in mind, there is no need to admit, e.g., worlds with only one or two objects or entirely symmetrical worlds as physically possible worlds (and see \citealp{Hacking:1975aa}, and \citealp{Belot:2001aa}, for an argument not to admit these as metaphysically possible worlds either). Furthermore, there is no need to abandon absolute discernibility in quantum physics either (since Bohmian mechanics solves the measurement problem by, among other things, respecting the individuality of the quantum particles; we will briefly mention Bohmian mechanics at the end of this paper). In a nutshell, its position in the network of distance relations distinguishes each matter point from all the other ones (absolute discernibility), but there is no fact as to which matter point has the position in question in that network (permutation invariance in the sense that labelling the matter points has no significance).

Since all there is to the matter points are the distance relations in which they stand, this is, like Cartesianism, a geometrical conception of matter. However, it is not to be confused with super-substantivalism, that is, the view that space (or space-time) is the only substance and matter a property of space. The main problem for this view is to account for motion, if there are only points of space (or space-time) and their topological and metrical properties, since these cannot move. Indeed, \cite{Wheeler:1962aa} tried super-substantivalism out in his programme of geometrodynamics, but failed in the attempt to reduce dynamical parameters to geometrical properties of points of space-time (see \citealp{Misner:1973aa}, § 44.3-4, in particular p. 1205). By contrast, if there are no points of space or space-time, but only distance relations between sparse points that hence are matter points, all the dynamical parameters that figure in physical theories can then be construed in terms of the role that they play in accounting for the change in these relations, that is, the motion of these points. In short, in a geometrical conception of matter by distance relations between sparse points, there is a clear sense in which there is motion and dynamical parameters capturing motion.

However, the substantivalist who accepts a dualism of matter and space can retort that by endorsing an absolute space that underlies the spatial configuration of matter, the substantivalist ontology, although being less simple than the relationalist one, gains in explanatory value. Thus, \citealp{Maudlin:2007aa}, pp. 87-89, takes length of a path in space as the primitive notion and derives the notion of distance of point particles from that notion, claiming that he is thus able to explain the constraints on the distance relation (such as the triangle inequality). But there is no gain in explanation here: the relationalist endorses certain relations as primitive. These relations exhibit constraints such as satisfying the triangle inequality. Consequently, in virtue of these relations being subject to these constraints, the world is one of matter points connected by distance relations (by contrast to e.g. a Cartesian world of mind points connected by thinking relations).

The substantivalist, to the contrary, traces these distances back to an underlying space. However, this space is construed as the space such that these constraints are satisfied: it comes with a metric in terms of, for example, paths of geodesic motion. Any metric defining a physical space is such that it fulfills all the constraints of three dimensional geometry. Hence, there is no additional explanatory value here in comparison to the relationalist who just presupposes that the relations admitted as primitive fulfill certain constraints; there only is the disadvantage that substantival space contains more structure than is needed to account for the experimental evidence, which consists in relative particle positions and change of these positions. In a nutshell, the substantivalist creates the illusion of giving a deeper explanation of something that, in fact, comes in a package with the postulation of a substantival space.

By the same token, the relationalist ontology stated in terms of the two axioms above accepts the whole change in the configuration of matter as primitive -- that is, the entire evolution of the distances among the matter points throughout the history of the universe. Again, this is no loss in explanation. Retracing this change to properties of the particles such as mass or charge, to forces or fields or wave functions, etc. does not provide a deeper explanation, since these are dynamical parameters that are \emph{defined} through the causal role that they play for the evolution of the particle configuration. Thus, one does not give a deeper explanation of attractive particle motion in terms of mass or the gravitational force, because these are defined through the effect that they have (or can have or are the power to have) on the motion of the particles. Taking explanations to end in the distance relations among matter points and their change endows this relationalist ontology with all the explanatory value that one can reasonably demand, namely to explain all the other phenomena in terms of the fundamental physical entities.

Hence, the argument against admitting dynamical parameters over and above change in distance relations to the ontology is the same as the argument against absolute space: doing so amounts to a commitment to a surplus structure that does not yield an additional explanatory value. Quite to the contrary, it leads to new drawbacks: in the case of absolute space, the commitment to a dualism of matter and space results in the impasse of not being able to come up with a cogent answer to the question of what it is that fills space. In the case of the dynamical parameters, it results in having to answer questions such as how a particle can reach out to other particles and change their motion in virtue of properties that are intrinsic to it (e.g. mass) or how a wave function, being defined on configuration space instead of being an entity in physical space, can influence the motion of matter in physical space. The argument for axiom \ref{a1} as well as axiom \ref{a2} boils therefore down to this one: bringing in more than what is admitted as primitive in these axioms not only amounts to a commitment to surplus structure, but this surplus structure also introduces new drawbacks, like for example the already mentioned Leibnizian arguments for ontological differences that make no physically observable difference.  

Let us now consider the second axiom. If there are only matter points connected by distances, all change is change in the distance relations among the permanent matter points. The change in the distance relations (and, hence, the motion of the matter points) can be represented as a parametrized list of states of the configuration of matter, that is, a map
\begin{align}
    \label{chng}
    \Delta_{(\cdot)}:\mathbb R \to \widetilde\Omega,
    \qquad
    \lambda\mapsto[\Delta_\lambda]_{\simeq},   
\end{align}
which, by means of $\lambda\mapsto[\Delta_\lambda]_{\simeq}$, denotes the change of the distance relations in the configuration of matter independently of a labelling. When formulating a dynamics on this basis, we have to assume that, following axiom \ref{a1}, the change takes place in such a way that the configuration cannot evolve into a state that violates requirements (\ref{one}) to (\ref{four}) in definition \ref{def:dist}: the distance relations individuate the matter points in any given state of the configuration, and the identity of the matter points across different states of the configuration is provided by their continuous trajectories. Note that, in order for the dynamics \eqref{chng} to make sense of the notion of particle trajectory, particles \emph{have to be} impenetrable and distinguishable (see \citealp[][section 1.2.3.]{Bach:1997aa}, for a rigorous proof of this statement).

As this minimalist ontology does not imply absolutism about space, in the same vein it does not imply absolutism about time: time derives from change. Again, one can follow Leibniz for whom time is the order of succession (see notably Leibniz' third letter, § 4, and fourth letter, § 41 in \citealp{Leibniz:1890aa}, pp. 363, 376). Hence, there is no time without change; but change exhibits an order, and what makes this order temporal is that it is unique and has a direction.

Although Leibnizian relationalism thus implies that the topology of time induced by the unique ordering of the elements in $\widetilde \Omega$ is absolute, there is no external measure of time: the idea that the global
dynamics unfolds according to the ticking of a universal clock is meaningless. If the entire universe could evolve at different external time rates, then two such evolutions would be physically indistinguishable, given that they would consist in the very same sequence of states of the universal configuration of spatial relations. Hence, they would exhibit the very same change in the distances among the matter points. In short, a commitment to an universal external clock would introduce ontological differences that would not make any physical difference (this point is made also in e.g. \citealp{Barbour:1982aa}, see especially pp. 296-297).
Consequently, there is no absolute metric of time. What we call ``time'' in this
context is just an arbitrary parametrization of the curve
$\lambda\mapsto[\Delta_\lambda]_{\simeq}$ on $\widetilde \Omega$
and not, as in the Newtonian case, an additional external variable. For this
reason, the only meaningful way to define a clock is to choose a reference
subsystem within the universe relative to which time is measured. An example of a simple reference subsystem is the circular motion of a pointer on a
dial of a watch, the arc length drawn by the pointer being directly related to
the parameter $\lambda$ in the definition of the map $\lambda\mapsto[\Delta_\lambda]_{\simeq}$.

Taking the topology of time to be absolute by no means contradicts relationalism, since time depends on the change in the configuration of matter points. This is in line with Mach's idea that ``time is an abstraction, at which we arrive by means of the change of things'' \citep[][p. 224]{Mach:1919aa}. Moreover, endorsing such a unique and directed order allows this relationalism to be compatible with both the $A$-series and the $B$-series view of time. On the latter, there is an objective, directed sequence of states of the configuration of matter so that the change in the configuration is ordered according to ``earlier'' and ``later''; but there is no past, present and future. The $A$-series view designates one state of the configuration as the present one. By the same token, this relationalism is compatible with both eternalism and presentism: one may take the whole stack of states of the configuration of matter to exist, or, endorsing the $A$-series view of time, maintain that only the present state of the configuration exists.

The commitment to a unique order of the change in the universal configuration of matter is a necessary condition also for the weaker, minimalist $C$-series view of time that only imposes an order on the change in the configuration of matter, but no direction (see \citealp{McTaggart:1908aa}, in particular pp. 458, 461-462). Hence, one may relax the Leibnizian relationalism about time by abandoning the requirement that the order in the change of the universal configuration of matter has a direction. The direction of time may not originate in that order as such, but in a particular initial configuration of matter (such as e.g. a low entropy initial configuration). However, if one also abandons the criterion of that order being unique, one arguably ends up in rejecting any notion of time, since not even the $C$-series can be recovered in this case. Such a radical stance is sometimes taken to be required for a proper understanding of time and change in general relativity (especially in its Hamiltonian formulation, as argued by \citealp{Earman:2002}; but see \citealp{Maudlin:2002aa} against \citealp{Earman:2002}). It is notably defended by \cite{Rovelli:2004aa} (in particular sections 1.3.1, 2.4, 3.2.4) in the context of quantum gravity (but see \citealp{Gryb:2015aa}, in particular pp. 23-24, 27-29, 35-38, for an argument that the relationalist stance to which we subscribe also covers the domain of -- quantum -- gravity).

\section{From ontology to physics: two strategies}
\label{sec:physics}

Although our concern here is not with necessary connections, we can formulate the minmalist ontology advocated in the preceding section in terms of Humeanism: the distance relations among points, making that these are matter points, and their change are the Humean mosaic. Everything else supervenes on that mosaic in the sense that, as far as physics is concerned, everything else comes in as the means to achieve the description of the distance relations among the matter points and their change throughout the entire history of the universe that strikes the best balance between simplicity and informational content. The argument for this Humean stance is its parsimony together with its empirical adequacy: less than distance relations individuating sparse points that then are matter points and the change of these relations would not do for an ontology of the physical world. Bringing in more is not only not necessary, but would also create new drawbacks instead of providing additional explanatory value. 

Putting this ontology in terms of Humeanism paves the way for introducing the simplest strategy to link it up with physics. This strategy consists in adopting physical theories as they stand and interpreting them as being committed to no more than this parsimonious ontology. When it comes to describing the distance relations and their change, it is appropriate to represent them as being embedded in a geometrical space (such as Euclidean space) and to introduce dynamical parameters that are constant (such as e.g. mass, charge, total energy, constants of nature) or that have an initial value (such as e.g. momenta, forces, fields, a wave function). This is appropriate because a physical theory seeks for dynamical laws that are such that by specifying an initial configuration of matter and putting that configuration into the law, all the -- past and future -- change in the configuration of matter is fixed. However, there is nothing about the mere distance relations in a given configuration of matter that yields such a law. In other words, there is nothing in a given configuration that fixes the -- past and future -- development of that configuration. That is why when seeking for a dynamical law, one usually attributes further parameters -- both geometrical and dynamical ones, over and above relative distances that change -- to the configuration of matter. However, the minimalist maintains that these parameters are nothing that pertains to the configuration of matter in addition to the distance relations; considering the evolution that these relations take throughout the history of the universe, these parameters are introduced when seeking for a law that describes that history in a simple and informative manner. In brief, whatever geometry and dynamics a physical theory employs, all this apparatus is there only as the means to achieve the description of the change in the distance relations that strikes the best balance between being simple and being informative. Consequently, this apparatus does not introduce ontological commitments that go beyond distance relations individuating matter points and their change.

In general, simplicity in ontology and simplicity in representation pull in opposite directions. Using only the concepts that describe what there is on the simplest ontology (matter points individuated by distance relations), the description of the evolution of the configuration of matter would not be simple at all, since one could only dress an extremely long list that enumerates all the change. Reading one's ontological commitments off from the simplest description -- such as e.g. Newtonian mechanics --, the ontology would not be simple at all: it would in this case be committed to absolute space and time, to momenta, gravitational masses, forces, etc. Humeanism allows us to have the best of these two worlds: simplicity in ontology achieved through parsimony and simplicity in description achieved through buying into the simplest physical theory that is empirically adequate.

\cite{Huggett:2006aa} shows how one can understand Euclidean geometry and Newtonian mechanics in a package as the Humean best system for a world of classical mechanics that consists only in distances among point particles and their change (see \citealp{Belot:2011aa}, pp. 60-77, for a criticism that spells out how the general objections against Humeanism apply in this case). The idea is that if one considers the entire history of the change in the distance relations in the configuration of matter points of the universe, the spatio-temporal geometry best suited to describe the universe is fixed together with the dynamical laws by the history of these relations as a whole. Given the fact that the change in the distance relations manifests certain salient patterns, Huggett singles out the notion of inertial motion as the idea of a particularly regular and simple motion. He then ties the notion of reference frame to that of a material body at rest at the origin of the frame. This makes it possible to relate different frames by means of continuous spatially rigid transformations. An inertial frame can then be defined as the frame in which the dynamical laws that supervene on the history of relations for the entire universe hold. In this way, the notion of inertial motion neither requires a substantival affine structure that singles out straight trajectories nor an absolute external time to which the uniformity of motion should be referred. Rather, these two structures supervene on purely relational facts. By the same token, absolute acceleration is reduced to the history of change of the spatial relations holding between an inertial and a non-inertial frame. Similarly, the regularities in the history of relations make it that Euclidean geometry is the simplest and most informative geometry representing that history. Such a framework clearly vindicates Leibnizian relationalism about space (spatial relations and their change are the ontological bedrock) and time (temporal facts supervene on the history, i.e. an ordered sequence, of instantaneous distance relations).

In the same vein, \citet[][§ 5.2]{Hall:2009aa} sketches out how dynamical parameters such as both inertial and gravitational mass as well as charge can be introduced as variables that figure in the laws of classical mechanics achieving the simplest and most informative description of the change in the relative particle positions throughout the history of the universe. Furthermore, this strategy has recently been applied to the quantum theories that admit a primitive ontology of matter in physical space in order to solve the measurement problem, such as Bohmian mechanics: in a nutshell, the universal wave function only has a nomological role, being a variable in the law that achieves the simplest and most informative description of the change in the primitive ontology (e.g. relative particle positions) throughout the history of the universe (see \citealp{Miller:2014aa}, \citealp{Esfeld:2014aa}, \citealp{Callender:2014aa}, \citealp{Bhogal:2015aa}). The availability of this Humean strategy shows that there is no point in reading off one's ontological commitments from the formalism of physical theories: one can be a scientific realist and yet be committed only to distance relations individuating matter points and the change of these relations, whatever other variables may figure in a physical theory that captures that change.

To mention a particularly striking example of this fact, consider a model of Newtonian mechanics with an angular momentum of the universe $\mathbf{J}$ that is greater than zero in the centre-of-mass rest frame. Obviously, a rotating universe is not conceivable in a relationalist ontology that admits only distance relations among matter points, but no space in which the configuration of matter is embedded. However, also in a Newtonian ontology of a universe rotating in absolute space, the rotation of the universe would manifest itself in certain changes in the distance relations among the point particles (e.g. in inhomogeneities in the cosmic microwave background with respect to our point of view). Consequently, the relationalist is free to interpret a value of angular momentum of the universe that is greater than zero as a convenient means to capture those changes in a simple and informative manner, without being committed to a space in which the universe rotates: describing those changes by using only variables for the change of relative distances would lead to a law of motion that is extremely complicated. Introducing further variables -- such as an angular momentum of the universe that can have a value greater than zero --, by contrast, would enable the formulation of a law of motion that is simple and elegant, but that is there only to capture the change in the relative distances among the point particles. In this manner, the relationalist can handle all kinds of bucket-like challenges.

Nonetheless, this strategy cannot recognize all the possible mathematical solutions of the dynamical equations of a physical theory as describing physically possible situations. As already mentioned at the end of the preceding section when discussing time, axiom \ref{a1} and requirement (\ref{four}) of definition \ref{def:dist} pose also a constraint on the dynamics: for instance, an evolution of the distance relations among the matter points that ends up in an entirely symmetrical configuration of the matter points is excluded in the same way as is a symmetrical initial configuration. Such solutions are a mathematical surplus of the formalism; the corresponding points in configuration space do not represent physically possible configurations of matter points. As argued in the preceding section, this is no objectionable restriction: having empirical adequacy in mind, there is no need to admit e.g. entirely symmetrical worlds as physically possible worlds.

This Humean strategy is a purely philosophical one: it is always available to vindicate the minimalist ontology set out in the previous section, physics be as it may, as long as all the evidence in fundamental physics comes down to relative particle positions and their change. In a nutshell, on this strategy, \emph{there is a relationalist ontology, but a non-relationalist physical theory}. This is no problem, since the non-relationalist theory can be interpreted in a cogent manner that is consistent with scientific realism as being committed to no more than the parsimonious relationalist ontology.

However, one may wonder whether buying into all the formal apparatus of, say, Newtonian mechanics is necessary to achieve a description of the change in the distance relations that meets the standards of simplicity and informational content. This reflection opens up the way for another strategy that can be dubbed \emph{alternative theory strategy}: instead of endorsing physical theories as they stand -- such as Newtonian mechanics -- and refusing to read ontological commitments off from their formalisms, one constructs alternative physical theories whose formal apparatus stays as close as possible to the ontology of there being only distance relations among point particles and their change and that matches the standard theories in their testable empirical predictions. In a nutshell, this strategy consists in \emph{building a relationalist physical theory on a relationalist ontology}.  

It is important to be clear about what the alternative theory strategy can achieve and what it cannot achieve: even if the ontology is exhausted by distance relations individuating matter points and the change of these relations, when it comes to formulating a dynamical law capturing that change, further dynamical parameters have to be introduced for the reason given above. Consequently, the alternative theory strategy has to admit dynamical parameters such as mass, constants of nature, initial momenta, etc. and cannot but resort to the Humean strategy in order to ban these parameters from the ontology. However, when it comes to space and time, the aim of this strategy is to avoid quantities that are tied to absolute space and time, such as empty space-time points, absolute velocities, absolute accelerations, or absolute rotations. As regards classical mechanics, the alternative theory strategy goes back at least to \cite{Mach:1919aa}. In the last three decades, it has been quite exhaustively worked out by Julian Barbour and collaborators (see \citealp{Barbour:1982aa}, for the seminal paper that laid down the ``best-matching'' framework). Furthermore, \citet{Belot:1999aa} and \citet{Saunders:2013aa} have also proposed each a relationalist theory of classical mechanics.

Let us discuss Belot's proposal first (see also \citealp{Pooley:2002aa}, section 5, for an appraisal of the framework). Belot starts by considering the Hamiltonian formulation of classical mechanics given in the language of symplectic geometry. Simply speaking, if we take $\mathcal{Q}$ to be the configuration space of a given system, then the cotangent bundle $T^{*}\mathcal{Q}$ is the phase space of the system (see \citealp[][section 2.3c]{Frankel:1997aa}, for technical details). This space comes equipped with a smooth function $H$, called the Hamiltonian, which -- roughly -- represents the total energy of the system, and a 2-form $\omega$, called the symplectic form. Glossing over the technical aspects, $\omega$ renders it possible to define a map $H\mapsto X_{H}$ that associates to the Hamiltonian a smooth vector field $X_{H}$ over  $T^{*}\mathcal{Q}$: such a map is nothing but an intrinsic representation of the usual Hamilton's equations. Hence, by integrating $X_{H}$ given some initial conditions $(\mathbf{q}_{0},\mathbf{\dot{q}}_{0})$, we get the unique curve in $T^{*}\mathcal{Q}$ that represents the dynamical evolution of the system under scrutiny. In the case of $N$ gravitating particles, $T^{*}\mathcal{Q}$ will be nothing but $\mathbb{R}^{6N}$. The justification of this fact is straightforward, if we consider what it takes to determine an initial condition $(\mathbf{q}_{0},\mathbf{\dot{q}}_{0})$: for each particle, we have to specify three numbers that give its position and further three numbers that specify the velocity vector ``attached'' to it. We immediately see in what sense this framework naturally fits a substantivalist understanding of space: two $N$-particle states that agree on all relational facts about the configuration (not only the relative positions, but also the relative orientations of the velocities), but disagree on how such a configuration is embedded in Euclidean space, would count as physically distinct possibilities. Then, the natural relationalist move would be to construct a relational configuration space $\mathcal{Q}_{0}$ by quotienting out from $\mathcal{Q}$ all the degrees of freedom associated with an embedding in Euclidean space, such as rigid translations and rotations. If we call $E(3)$ the set of isometries of Euclidean $3$-space, then $\mathcal{Q}_{0}=\mathcal{Q}/E(3)$: this construction assures us that distinct points in $\mathcal{Q}$ that represent the same relational configuration ``collapse'' to the same point in $\mathcal{Q}_{0}$. Note that (i) $\mathcal{Q}_{0}$ admits a well-defined cotangent bundle $T^{*}\mathcal{Q}_{0}$, which is equipped with a well-behaved symplectic structure, and (ii) that the starting Hamiltonian defined on $\mathcal{Q}$ admits a smooth projection $H_{0}$ to $T^{*}\mathcal{Q}_{0}$ because it is invariant under the action of $E(3)$.

Belot's theory qualifies as relational, since the ontological facts making up a set of initial data do not encode any notion of position in absolute space or absolute velocity, and the laws of motion specify how these initial data evolve; furthermore, the dynamical laws of the theory are fully defined on relational phase space. However, there are at least three concerns that one can raise about Belot's proposal. In the first place, this theory still has a notion of absolute time inherent in the dynamical laws. In fact, there is nothing in the quotienting out procedure that leads from a dynamics over $\mathcal{Q}$ to one over $\mathcal{Q}_0$ that eliminates the absolute temporal metric of Newtonian mechanics, which means that the very same succession of purely relational configurations can unfold at different rates depending on the ticking of an universal external clock. Secondly, there is a clear sense in which spatial relations are Euclidean from the beginning: they are just equivalence classes of embedding degrees of freedom as encoded in $E(3)$. Thirdly, Belot's theory, despite being very close to Newtonian mechanics, is not as empirically predictive as its absolute counterpart. This is obvious because, if we think about all the initial data that are needed in the Newtonian theory, we realize that they must include the rate of change in the orientation of the configuration of $N$ particles with respect to absolute space: in the passage from $\mathcal{Q}$ to $\mathcal{Q}_{0}$, this information is simply washed away. In particular, the Newtonian theory admits models with non vanishing total angular momentum $\mathbf{J}$ of the universe. We repeat that this is not just a metaphysical aspect, but a physical one in the sense that the condition $\mathbf{J}\neq 0$ carries with it empirically testable consequences. Belot's proposal to overcome the problem is just to bite the bullet: his relational reduction indeed recovers only a part of the Newtonian one, but -- given that up to now we have reliable experimental evidence that the universe is not rotating -- it recovers exactly the empirically adequate part.

Let us turn to Barbour's proposal (see \citealp{Barbour:1982aa,Barbour:2003aa,Barbour:2012aa} for the original resources; \citealp{Pooley:2002aa}, sections 6-7, give an excellent overview of the framework, together with some cogent philosophical considerations). Barbour's relationalist motivations are the same as Belot's, that is, to eliminate all the spatial degrees of freedom that produce no observable difference. However, Barbour extends this requirement to temporal degrees of freedom as well. By way of consequence, the construction of his framework involves two steps, namely the implementation of (i) spatial and (ii) temporal relationalism. As regards the first step, Barbour adopts the same strategy as Belot: he takes standard configuration space $\mathcal{Q}$ and quotients out all Euclidean isometries, comprised of scale transformations, which means that he quotients out also the degrees of freedom related to ``stretchings'' or ``shrinkings'' of configurations that preserve the ratio of distances. This means that he considers a wider group than $E(3)$, namely the similarity group $Sim(3)$. Hence, his relational configuration space $\mathcal{Q}_{0}=\mathcal{Q}/Sim(3)$ is aptly called \emph{shape} space, because each configuration in there is individuated by its form and not by its size.

The second step is technically more complicated: firstly he defines an ``intrinsic'' difference that measures how similar two shapes are. This difference is expressed in terms of ``best-matching'' coordinates. Intuitively, we imagine the two shapes laid down over two distinct Cartesian coordinate grids $O$ and $O'$; then we hold fixed the first shape and grid and ``move'' the second by applying transformations in $Sim(3)$ until the two shapes are juxtaposed as close as possible. The best-matching coordinates are then defined as the overlap deficit $O-O'$ between the two coordinate grids. Secondly, he uses this intrinsic metric to define a Jacobi action, thus setting a Jacobi variational principle on $\mathcal{Q}_{0}$ (see \citealp{Lanczos:1970}, pp. 132-140, for a technical introduction to the Jacobi principle in classical mechanics). The Jacobi action is reparametrization invariant, that is, it does not change whatever ``time'' parameter we choose.

With this machinery in place, carrying out the variation of the action with respect to the best-matching coordinates, we obtain a set of generalized Euler-Lagrange equations whose integral curves are nothing but the geodesics of $\mathcal{Q}_{0}$. Given some initial conditions, one of these curves is singled out, which represents the dynamical evolution of the system. This evolution is given in fully relationalist terms: the curve singled out by the equations plus the initial conditions represents a list of relational configurations, which is parametrized by an arbitrary monotonically increasing parameter: hence, there is no external clock that measures dynamical change; on the contrary, it is the change in the list of configurations that enables an (arbitrary) parametrization. The important point is that there exists a particular parametrization of the curve for which the generalized Euler-Lagrange equations take the usual Newtonian form. Thus, if we adopt this (again, arbitrary) parametrization, we obtain a dynamical description that matches the Newtonian one. In this sense, Newtonian mechanics comes out of Barbour's framework by means of something closely resembling a gauge fixing. The descriptive simplicity of the Newtonian formulation then explains why, historically, classical physics was framed in these terms.

There are at least three critical points about this framework worth being highlighted. Firstly, no usual Newtonian potential is compatible with the condition of scale-invariance. Even if it is always possible to reproduce the form of the most usual classical potentials in the appropriate gauge by a clever mathematical manoeuvre, still this mimicking strategy might lead to unwanted physical restrictions, such as no angular momentum exchange between subsystems (see \citealp[][section 5.1.2]{Anderson:2013aa}, for a technical discussion of this point). Secondly, the implementation of a geodesic principle on a general shape space might not always be that straightforward. In general cases, in fact, the quotienting out procedure sketched above leads to a shape space whose global geometry is that of a stratified manifold, where each ``stratum'' is a sub-manifold that can differ from the others in many respects, including the dimensionality. It is then quite intuitive to understand that, if $\mathcal{Q}_{0}$ is a stratified manifold, it is problematic to account for a dynamical evolution given in terms of a geodesic trajectory that hits different strata of $\mathcal{Q}_{0}$ (see \citealp[][section 9.4 and references mentioned therein]{Anderson:2015aa}, for discussion). The moral is that Barbour's framework works well in a suitably small region of $\mathcal{Q}_{0}$, but might break down on a larger scale, depending on the particular geometrical structure of $\mathcal{Q}_{0}$. Thirdly, as a result of quotienting out the group of rotations from $\mathcal{Q}$, one gets the condition $\mathbf{J}=0$ as a constraint on shape space dynamics. However, even if the actual universe satisfies the condition of $\mathbf{J}=0$, it is desirable that a relationalist theory should be able, as far as possible, to accommodate observable consequences ascribable in absolute terms to a non-vanishing total angular momentum of the universal configuration of matter.

Finally, let us consider the proposal spelled out in \citet{Saunders:2013aa}. Like Belot and Barbour, Saunders' aim is to dispense with absolute quantities of motion. However, unlike the former two, he also seeks to save the core conceptual structure of Newton's \emph{Principia}. In order to do so, he shows that the Newtonian laws can be cast in terms of directed distances representing inter-particle separations. This is possible because the absolute notion of ``straight trajectory'' needed to make sense of inertial motion -- which, in turn, is required for rendering Newton's first and second law meaningful --  involves too much structure, namely, a privileged affine connection (that of neo-Newtonian space-time). Instead, Newton's laws can make perfect sense even if we replace the talk of straight trajectories with that of relative velocities not changing over time, and this can be accounted for not just by a single preferred connection, but by a whole class of affine connections whose time-like geodesics are mutually non-rotating. This is all that Saunders needs in order to account for accelerations and rotations in relationalist terms: a space-time manifold equipped with enough structure to allow for the comparison of spatial directions (and related angles) at different times (what Saunders calls ``Newton-Huygens'' space-time, see also \citealp{Earman:1989aa}, pp. 31-32, for a formal characterization of this space-time). In the Newton-Huygens space-time, unlike the neo-Newtonian one, it is meaningless to talk about the absolute acceleration of a particle, or even its inertial motion (two notions that are tied to a privileged affine connection), while it is perfectly meaningful to ask questions about the change of orientation of a configuration in time.

The huge virtue of this framework is that it is able to recover the full spectrum of Newtonian models, thus accounting also for $\mathbf{J}\neq0$ cases, without invoking absolute notions. Given, in fact, that differences in direction can be defined relationally by admitting a primitive notion of parallelism, and then defining change in direction by comparison of spatial relations at different times, Saunders' theory can account for global rotations in terms of relational quantities. However, the fact that this theory makes it meaningful to compare spatial directions at different times represents a substantial weakening of the relationalist programme. This framework is much less relationalist than Belot's and Barbour's, for which the excision of any physical meaning attached to global rotations is a constitutive feature. In this sense, Saunders' theory is a ``halfway house'' form of a relationalism, as he himself notes (\citealp{Saunders:2013aa}, p. 44).

In sum, there are well-grounded reservations whether these relationalist theories fully implement a relationalist ontology. That ontology is relationalist both with respect to space and with respect to time, whereas Belot's and Saunders' theories are relationalist only with respect to space. Furthermore, that ontology is not tied to a particular geometry such as Euclidean geometry: a configuration of matter points satisfying axiom \ref{a1} and definition \ref{def:dist} does not carry with it any primitive geometrical fact that singles out a distinguished space in which it has to be embedded. By contrast, there is a clear sense in which Belot's and Barbour's spatial relations are Euclidean from the outset: each point in $\mathcal{Q}_0$ can be seen as an equivalence class of Euclidean configurations; there is no way, by fixing a certain gauge, to end up with a configuration embedded in a non-Euclidean space. Also Saunders' relations are inherently Euclidean, since Newton-Huygens space-time encodes the structure of a series of instantaneous $3$-dimensional affine spaces equipped with an Euclidean metric. Even worse, all these theories rely on more primitive structure than just distances. The very concept of shape requires primitive facts about angles to be meaningful, so that Barbour's ontology has to include a conformal structure. Saunders' ontology requires not only distances, but \emph{directed} distances, which means that some primitive geometrical facts have to be postulated (especially those making up a standard of space-like parallel transport), which are encoded in an affine structure.  

At this point, one may legitimately ask whether it is possible to resort to the Humean strategy to argue that the additional structure of these theories is just part of the package we get when seeking for the best description of motion. The answer is that such a move, in this case, raises substantial worries. Put simply, a Humean justification of the surplus structure would imply that what these theories do is basically to ``embellish'' a set of relational initial data $\Delta_0$ by embedding them in a more structured set $Q_0$ and then using the equations of motion cast in terms of this surplus structure to evolve these data until reaching the result $Q_t$, from which the relational solution $\Delta_\lambda$ would be read off. But that would have unwanted implications. The first of these is that, in this way, neither of the above theories could be considered even mildly relationalist anymore, being parasitic on a dynamical description that involves an irreducible surplus structure. In such a case, it would be awkward to prefer these mathematically elaborated theories to Newtonian gravitation, given that Huggett's Humean strategy works perfectly for the purposes of a relationalist ontology in that context. The second implication is that the strategy of evolving relational data by ``stealing a ride'' to a non-relationalist dynamics and then discarding the surplus structure as a mere representational means would suspiciously look like a trivial instrumentalist move, as discussed, e.g., by \citet[][p. 128]{Earman:1989aa} and \citet[][p. 10]{Belot2000:aa}.

In short, combining a minimalist relationalist ontology with the alternative theory strategy faces a dilemma: either one insists that angles or directions are just part of the Humean package, thus ending up with -- using Earman's words -- a ``cheap instrumentalist rip-off'' of a theory that in any case does not qualify as a genuine relationalist competitor to Newtonian mechanics; or one bites the bullet and introduces more primitive structure in the ontology (be it a conformal or an affine one), thus compromising the original motivations for relationalism from ontological parsimony.

Let us now briefly consider how the Humean and the alternative theory strategy fare when it comes to quantum mechanics. Adopting relationalism about space and time is a reasonable option if one pursues a solution to the quantum measurement problem in terms of being committed to what is known as a primitive ontology of matter being distributed in ordinary space-time (and not just the quantum state defined on configuration space). Bohmian mechanics is the most prominent primitive ontology theory of quantum mechanics, setting out a primitive ontology of point particles moving in a three-dimensional, Euclidean space (see \citealp{Durr:2013aa}). Given that the laws of Bohmian dynamics, although being different from the Newtonian ones, are nonetheless formulated over an absolute Newtonian spatio-temporal background, all the options for relationalism discussed in this section can be applied to Bohmian mechanics as well.

Concerning the strategy to interpret a non-relationalist theory as being committed only to a relationalist ontology, we already mentioned the Humean treatment of the wave function at the beginning of this section. Space-time in Bohmian mechanics poses no problem for Humeanism: the manner in which \cite{Huggett:2006aa} deals with space-time in the Newtonian case can simply be applied to the Bohmian case. As regards the alternative theory strategy, the proposal of \cite{Belot:1999aa} would require forcing the Bohmian formalism in a Hamiltonian context; this is possible, but results in a quite unreasonably complicated formalism with a huge amount of surplus descriptive structure, which is not needed by Bohmian dynamics (see \citealp{Holland:2001aa,Holland:2001ab} for a decently worked out Hamiltonian version of Bohmian mechanics). Barbour's framework, by contrast, can in a natural way be applied to Bohmian mechanics (see \citealp{Vassallo:2015aa}, \citealp{Vassallo:2016aa}). The same goes for the milder relationalist approach of \cite{Saunders:2013aa}, since this approach would basically amount to rewrite the Bohmian theory in a way that makes it meaningless to refer distances and directions to any point taken as ``origin''.

We cannot go into relativity physics in this paper for lack of space. We would just like to make two brief remarks concerning quantum field theory and general relativity theory, respectively. (i) The relationalist ontology set out in section 1 can be carried on to quantum field theory: this theory is hit by the measurement problem in the same way as quantum mechanics (see \citealp{Barrett:2014aa}). The Bohmian solution to the measurement problem can be applied to quantum field theory in the same way as to quantum mechanics, even in the form of an ontology of permanent point particles moving according to a deterministic law that explains the statistics of the appearances of particle creation and annihilation phenomena in the experiments (this approach is known as Dirac sea Bohmian quantum field theory; see \citealp{Colin:2007aa}, \citealp{Deckert:2010aa}, chs. 6-7). However, it is an open issue whether and how the relationalist approaches discussed in this section can be carried on from Bohmian mechanics to such a Bohmian quantum field theory.

(ii) When it comes to general relativity theory, also in this domain, as in any other field theory, fields are tested by the motion of particles. There is no direct evidence of fields. All the evidence is one of relative particle motion (cf. \citealp{Einstein:1949aa}, p. 209). Against this background, a Humean strategy of treating the formal apparatus of general relativity theory as being the means to achieve a description of the overall relative particle motion that strikes the best balance between being simple and being informative about that motion seems to be an option that would be worth trying out (see \citealp{Vassallo:2016ab}, for a concrete proposal in this sense). As regards the alternative theory strategy, Barbour and collaborators have developed an alternative theory of gravitation in the relativistic regime (see notably \citealp{Barbour:2002aa}, \citealp{Barbour:2012aa}), which has recently attracted attention in the philosophical literature (see e.g. \citealp{Gryb:2015aa}).

In conclusion, we have seen that, when vindicating a relationalist ontology for classical mechanics consisting of matter points individuated by distance relations and the change of these relations only, the Humean strategy of combining a relationalist ontology with a non-relationalist physical theory and interpreting that latter theory as being committed to no more than the minimalist relationalist ontology is coherent and always available as a fall back option for the relationalist. Pursuing the more ambitious strategy of developing a relationalist physical theory that is an alternative to the standard non-relationalist physics can, as things stand, not be carried out without compromising the relationalist ontology -- at least by including a primitive conformal structure, or even by including a primitive affine structure (when seeking to recover all models of Newtonian mechanics that may have observable consequences).

\paragraph{Acknowledgements.} We are grateful to Vincent Lam, Dustin Lazarovici, Andrea Oldofredi and Christian Wüthrich for helpful discussions. A. Vassallo's work on this paper was supported by the Swiss National Science Foundation, grant no. 105212\_149650, while D.-A.\ Deckert's work was funded by the junior research group grant \emph{Interaction between Light and Matter} of the Elite Network of Bavaria.

\bibliographystyle{apalike}
\bibliography{references_fundont}

\begin{thebibliography}{}

\bibitem[Anderson, 2013]{Anderson:2013aa}
Anderson, E. (2013).
\newblock The problem of time and quantum cosmology in the relational particle
  mechanics arena.
\newblock {\em arXiv:1111.1472v3 [gr-qc]}.
\newblock \url{http://arxiv.org/abs/1111.1472}.

\bibitem[Anderson, 2015]{Anderson:2015aa}
Anderson, E. (2015).
\newblock Configuration spaces in fundamental physics.
\newblock {\em arXiv:1503.01507v2 [gr-qc]}.
\newblock \url{http://arxiv.org/abs/1503.01507}.

\bibitem[Ariew, 2000]{Leibniz:2000aa}
Ariew, R., editor (2000).
\newblock {\em G. {W}. {L}eibniz and {S}. {C}larke: {C}orrespondence}.
\newblock Indianapolis: Hackett.

\bibitem[Bach, 1997]{Bach:1997aa}
Bach, A. (1997).
\newblock {\em Indistinguishable classical particles}.
\newblock Berlin: Springer.

\bibitem[Barbour, 2003]{Barbour:2003aa}
Barbour, J. (2003).
\newblock Scale-invariant gravity: particle dynamics.
\newblock {\em Classical and Quantum Gravity}, 20:1543--1570.

\bibitem[Barbour, 2012]{Barbour:2012aa}
Barbour, J. (2012).
\newblock Shape dynamics. {A}n introduction.
\newblock In Finster, F., M{\"u}ller, O., Nardmann, M., Tolksdorf, J., and
  Zeidler, E., editors, {\em Quantum field theory and gravity}, pages 257--297.
  Basel: Birkh{\"a}user.

\bibitem[Barbour and Bertotti, 1982]{Barbour:1982aa}
Barbour, J. and Bertotti, B. (1982).
\newblock Mach's principle and the structure of dynamical theories.
\newblock {\em Proceedings of the Royal Society A}, 382:295--306.

\bibitem[Barbour et~al., 2002]{Barbour:2002aa}
Barbour, J., Foster, B., and OMurchadha, N. (2002).
\newblock Relativity without relativity.
\newblock {\em Classical and Quantum Gravity}, 19:3217--3248.

\bibitem[Barrett, 2014]{Barrett:2014aa}
Barrett, J.~A. (2014).
\newblock Entanglement and disentanglement in relativistic quantum mechanics.
\newblock {\em Studies in History and Philosophy of Modern Physics},
  48:168--174.

\bibitem[Belot, 1999]{Belot:1999aa}
Belot, G. (1999).
\newblock Rehabilitating relationalism.
\newblock {\em International Studies in the Philosophy of Science}, 13:35--52.

\bibitem[Belot, 2000]{Belot2000:aa}
Belot, G. (2000).
\newblock Geometry and motion.
\newblock {\em British Journal for the Philosophy of Science}, 51(4):561--595.

\bibitem[Belot, 2001]{Belot:2001aa}
Belot, G. (2001).
\newblock The principle of sufficient reason.
\newblock {\em Journal of Philosophy}, 98:55--74.

\bibitem[Belot, 2011]{Belot:2011aa}
Belot, G. (2011).
\newblock {\em Geometric possibility}.
\newblock Oxford: Oxford University Press.

\bibitem[Bhogal and Perry, 2016]{Bhogal:2015aa}
Bhogal, H. and Perry, Z.~R. (2016).
\newblock What the {H}umean should say about entanglement.
\newblock {\em No{\^u}s}, page DOI 10.1111/nous.12095.

\bibitem[Blackburn, 1990]{Blackburn:1990aa}
Blackburn, S. (1990).
\newblock Filling in space.
\newblock {\em Analysis}, 50:62--65.

\bibitem[Callender, 2015]{Callender:2014aa}
Callender, C. (2015).
\newblock One world, one beable.
\newblock {\em Synthese}, 192(10):3153--3177.

\bibitem[Colin and Struyve, 2007]{Colin:2007aa}
Colin, S. and Struyve, W. (2007).
\newblock A {D}irac sea pilot-wave model for quantum field theory.
\newblock {\em Journal of Physics A}, 40(26):7309--7341.

\bibitem[Deckert, 2010]{Deckert:2010aa}
Deckert, D.-A. (2010).
\newblock {\em Electrodynamic absorber theory -- a mathematical study}.
\newblock T{\"o}nning: Der Andere Verlag.

\bibitem[D{\"u}rr et~al., 2013]{Durr:2013aa}
D{\"u}rr, D., Goldstein, S., and Zangh{\`\i}, N. (2013).
\newblock {\em Quantum physics without quantum philosophy}.
\newblock Berlin: Springer.

\bibitem[Earman, 1989]{Earman:1989aa}
Earman, J. (1989).
\newblock {\em World enough and space-time. Absolute versus relational theories
  of spacetime}.
\newblock Cambridge, Massachusetts: MIT Press.

\bibitem[Earman, 2002]{Earman:2002}
Earman, J. (2002).
\newblock Thoroughly modern {M}c{T}aggart or what {M}c{T}aggart would have said
  if he had read the general theory of relativity.
\newblock {\em Philosopher's Imprint}, 2(3).
\newblock \url{http://www.philosophersimprint.org/002003/}.

\bibitem[Einstein and Infeld, 1949]{Einstein:1949aa}
Einstein, A. and Infeld, L. (1949).
\newblock On the motion of particles in general relativity theory.
\newblock {\em Canadian Journal of Mathematics}, 1:209--241.

\bibitem[Esfeld, 2014]{Esfeld:2014aa}
Esfeld, M. (2014).
\newblock Quantum {H}umeanism, or: physicalism without properties.
\newblock {\em The Philosophical Quarterly}, 64(256):453--470.

\bibitem[Esfeld and Lam, 2011]{Esfeld:2011aa}
Esfeld, M. and Lam, V. (2011).
\newblock Ontic structural realism as a metaphysics of objects.
\newblock In A. and Bokulich, P., editors, {\em Scientific structuralism},
  pages 143--159. Dordrecht: Springer.

\bibitem[Frankel, 1997]{Frankel:1997aa}
Frankel, T. (1997).
\newblock {\em The geometry of physics}.
\newblock Cambridge: Cambridge University Press.

\bibitem[French, 2014]{French:2014aa}
French, S. (2014).
\newblock {\em The structure of the world. Metaphysics and representation}.
\newblock Oxford: Oxford University Press.

\bibitem[Gerhardt, 1890]{Leibniz:1890aa}
Gerhardt, C.~I., editor (1890).
\newblock {\em Die philosophischen {S}chriften von {G}. {W}. {L}eibniz. {B}and
  7}.
\newblock Berlin: Weidmannsche Verlagsbuchhandlung.

\bibitem[Gryb and Th{\'e}bault, 2016]{Gryb:2015aa}
Gryb, S. and Th{\'e}bault, K. P.~Y. (2016).
\newblock Time remains.
\newblock {\em British Journal for the Philosophy of Science}, page DOI
  10.1093/bjps/axv009.

\bibitem[Hacking, 1975]{Hacking:1975aa}
Hacking, I. (1975).
\newblock The identity of indiscernibles.
\newblock {\em Journal of Philosophy}, 72:249--256.

\bibitem[Hall, 2009]{Hall:2009aa}
Hall, N. (2009).
\newblock Humean reductionism about laws of nature.
\newblock Unpublished manuscript. http://philpapers.org/rec/halhra.

\bibitem[Holland, 2001a]{Holland:2001aa}
Holland, P. (2001a).
\newblock Hamiltonian theory of wave and particle in quantum mechanics {I}:
  {L}iouville's theorem and the interpretation of the de {B}roglie-{B}ohm
  theory.
\newblock {\em Il Nuovo Cimento B}, 116:1043--1070.

\bibitem[Holland, 2001b]{Holland:2001ab}
Holland, P. (2001b).
\newblock Hamiltonian theory of wave and particle in quantum mechanics {II}:
  {H}amilton-{J}acobi theory and particle back-reaction.
\newblock {\em Il Nuovo Cimento B}, 116:1143--1172.

\bibitem[Huggett, 2006]{Huggett:2006aa}
Huggett, N. (2006).
\newblock The regularity account of relational spacetime.
\newblock {\em Mind}, 115(457):41--73.

\bibitem[Ladyman, 2007]{Ladyman:2007aa}
Ladyman, J. (2007).
\newblock On the identity and diversity of objects in a structure.
\newblock {\em Proceedings of the Aristotelian Society. Supplementary Volume},
  81(1):23--43.

\bibitem[Ladyman and Ross, 2007]{Ladyman:2007b}
Ladyman, J. and Ross, D. (2007).
\newblock {\em Every thing must go: metaphysics naturalized}.
\newblock New York: Oxford University Press.

\bibitem[Lanczos, 1970]{Lanczos:1970}
Lanczos, C. (1970).
\newblock {\em The variational principles of mechanics}.
\newblock University of Toronto Press, fourth edition.

\bibitem[Lewis, 1986]{Lewis:1986ab}
Lewis, D. (1986).
\newblock {\em On the plurality of worlds}.
\newblock Oxford: Blackwell.

\bibitem[Locke, 1690]{Locke:1690aa}
Locke, J. (1690).
\newblock {\em An essay concerning human understanding}.

\bibitem[Mach, 1919]{Mach:1919aa}
Mach, E. (1919).
\newblock {\em The science of mechanics: a critical and historical account of
  its development. Fourth edition. Translation by Thomas J. McCormack}.
\newblock Chicago: Open Court.

\bibitem[Maudlin, 1993]{Maudlin:1993aa}
Maudlin, T. (1993).
\newblock Buckets of water and waves of space: why spacetime is probably a
  substance.
\newblock {\em Philosophy of Science}, 60:183--203.

\bibitem[Maudlin, 2002]{Maudlin:2002aa}
Maudlin, T. (2002).
\newblock Thoroughly muddled {M}c{T}aggart or how to abuse gauge freedom to
  create metaphysical monstrosities.
\newblock {\em Philosopher's Imprint}, 2(4).

\bibitem[Maudlin, 2007]{Maudlin:2007aa}
Maudlin, T. (2007).
\newblock {\em The metaphysics within physics}.
\newblock New York: Oxford University Press.

\bibitem[McTaggart, 1908]{McTaggart:1908aa}
McTaggart, J.~E. (1908).
\newblock The unreality of time.
\newblock {\em Mind}, 17(68):457--474.

\bibitem[Miller, 2014]{Miller:2014aa}
Miller, E. (2014).
\newblock Quantum entanglement, {B}ohmian mechanics, and {H}umean
  supervenience.
\newblock {\em Australasian Journal of Philosophy}, 92:567--583.

\bibitem[Misner et~al., 1973]{Misner:1973aa}
Misner, C.~W., Thorne, K.~S., and Wheeler, J.~A. (1973).
\newblock {\em Gravitation}.
\newblock San Francisco: Freeman.

\bibitem[Muller, 2011]{Muller:2011aa}
Muller, F.~A. (2011).
\newblock How to defeat {W}{\"u}thrich's abysmal embarassment argument against
  space-time structuralism.
\newblock {\em Philosophy of Science}, 78:1046--1057.

\bibitem[Pooley, 2013]{Pooley:2013aa}
Pooley, O. (2013).
\newblock Substantivalist and relationalist approaches to spacetime.
\newblock In Batterman, R., editor, {\em The {O}xford handbook of philosophy of
  physics}, pages 522--586. Oxford: Oxford University Press.

\bibitem[Pooley and Brown, 2002]{Pooley:2002aa}
Pooley, O. and Brown, H. (2002).
\newblock Relationalism rehabilitated? {I}: {C}lassical mechanics.
\newblock {\em British Journal for the Philosophy of Science}, 53:183--204.

\bibitem[Rovelli, 2004]{Rovelli:2004aa}
Rovelli, C. (2004).
\newblock {\em Quantum gravity}.
\newblock Cambridge: Cambridge University Press.

\bibitem[Saunders, 2006]{Saunders:2006aa}
Saunders, S. (2006).
\newblock Are quantum particles objects?
\newblock {\em Analysis}, 66:52--63.

\bibitem[Saunders, 2013]{Saunders:2013aa}
Saunders, S. (2013).
\newblock Rethinking {N}ewton's {P}rincipia.
\newblock {\em Philosophy of Science}, 80(1):22--48.

\bibitem[Vassallo, 2015]{Vassallo:2015aa}
Vassallo, A. (2015).
\newblock Can {B}ohmian mechanics be made background independent?
\newblock {\em Studies in History and Philosophy of Modern Physics},
  52:242--250.

\bibitem[Vassallo and Esfeld, 2016]{Vassallo:2016ab}
Vassallo, A. and Esfeld, M. (2016).
\newblock Leibnizian relationalism for general relativistic physics.
\newblock {\em Studies in History and Philosophy of Modern Physics}.
\newblock DOI 10.1016/j.shpsb.2016.08.006.

\bibitem[Vassallo and Ip, 2016]{Vassallo:2016aa}
Vassallo, A. and Ip, P.~H. (2016).
\newblock On the conceptual issues surrounding the notion of relational
  {B}ohmian dynamics.
\newblock {\em Foundations of Physics}, 46:943--972.

\bibitem[Wheeler, 1962]{Wheeler:1962aa}
Wheeler, J.~A. (1962).
\newblock {\em Geometrodynamics}.
\newblock New York: Academic Press.

\bibitem[W{\"u}thrich, 2009]{Wuthrich:2009aa}
W{\"u}thrich, C. (2009).
\newblock Challenging the spacetime structuralist.
\newblock {\em Philosophy of Science}, 76:1039--1051.

\end{thebibliography}

\end{document}